\documentclass[twocolumn,aps,showpacs,preprintnumbers,amsmath,amssymb,superscriptaddress]{revtex4-1}
\usepackage[dvips]{graphicx}
\usepackage{bm}
\usepackage[usenames,dvipsnames]{color}
\usepackage{amssymb}
\usepackage{siunitx}
\usepackage{framed}

\newcommand{\GCI}{GdCoIn$_5 $}
\newcommand{\GRI}{GdRhIn$_5 $}
\newcommand{\TN}{$T_N$~}
\begin{document}
\author{D. Betancourth}
\author{V. F. Correa}
\author{Jorge I. Facio}
\author{J. Fern\'andez}
\affiliation{Centro At\'omico Bariloche and Instituto Balseiro,  Comisi\'on Nacional de Energ\'ia At\'omica and CONICET, 8400 Bariloche, Argentina}
\author{V. Vildosola}
\affiliation{Departamento de Materia Condensada, GIyA, CNEA, CONICET, (1650) San Mart\'{\i}n, Provincia de Buenos Aires, Argentina}
\author{A. A. Aligia}
\author{Pablo S. Cornaglia}
\author{D. J. Garc\'ia}
\affiliation{Centro At\'omico Bariloche and Instituto Balseiro,  Comisi\'on Nacional de Energ\'ia At\'omica and CONICET, 8400 Bariloche, Argentina}

\title {Magnetostriction Reveals Orthorhombic Distortion in Tetrahedral Gd-compounds}
\begin{abstract}
    We report detailed thermal expansion and magnetostriction experiments on \GCI~and~\GRI~ single crystal samples that show a sudden change in the dilation at a field $B^\star$ for temperatures below the N\'eel transition temperature $T_N$. We present a first-principles model including crystal-field effects, dipolar and exchange interactions, and the dependence of the exchange couplings with lattice distortions in order to fully account for the magnetostriction and magnetic susceptibility data. 
The mean-field solution of the model shows that a transition between metastable states occurs at the field $B^\star$.
It also indicates that two degenerate phases coexist in the sample at temperatures below $T_N$. 
This allows to explain the lack of observation, in high resolution x-ray experiments, of an orthorhombic distortion at the N\'eel transition even though the magnetic structure breaks the tetragonal symmetry and the magnetoelastic coupling is significant. These conclusions could be extended to other tetragonal Gd-based compounds that present the same phenomenology.
\end{abstract}
\maketitle


\section{Introduction}
Rare-earth magnetic compounds are among the strongest permanent magnets and present the highest magnetostrictive responses ever recorded. These remarkable properties stem from the large magnetic moments of the rare-earth ions with partially filled f-shells and the magnetic anisotropy associated with crystal-field effects and spin-orbit couplings. The magnetic structure of these compounds is mainly determined by the Ruderman-Kittel-Kasuya-Yosida (RKKY) exchange interactions between the magnetic moments of the rare earth ions and by the crystal field, which dominate over the dipolar interaction.

In many materials the magnetoelastic couplings lead to a significant spontaneous lattice distortion concomitant with the magnetic order at zero applied magnetic field. In some systems, the associated changes in the lattice constants can be as large as a few percent~\cite{rotter2002spontaneous}. When the symmetry of the magnetic order is lower than the lattice symmetry, a reduction of the latter is expected at the magnetic transition. This can occur, e.g., when the magnetoelastic couplings generate changes in the lattice parameters which do not preserve the lattice symmetry.

In the rare earth series, Gadolinium stands as different. In solid state compounds it is generally found as a trivalent ion Gd$^{3+}$ for which Hund's rules indicate the maximum spin allowed $\mathcal{S}=7/2$, and zero angular momentum $\mathcal{L}=0$. 
 As a consequence of the latter, the magnetic coupling to the lattice via crystal-field effects is expected to be weak. The magnetic anisotropy observed in Gd compounds is therefore usually attributed to the dipolar interaction~\cite{rotter2003dipole}. The dependence of the exchange couplings on the relative distance between ions can, however, give rise to large magnetoelastic couplings. 

Although a coupling between lattice and magnetic orders is observed on most Gd compounds there are few reports of lattice symmetry breaking at the magnetic transition (see Ref. \cite{rotter2002spontaneous} for a review). To the best of our knowledge, such lattice symmetry breaking in Gd compounds has only been confirmed for the ferromagnetic GdZn compound where the lattice distorts from cubic to orthorhombic at the magnetic transition~\cite{rouchy1981magnetic}. 

Many Gd compounds with tetragonal lattice show antiferromagnetic (AFM) order \cite{rotter2003dipole,lindbaum2002chapter}. Competing magnetic couplings can result in an AFM order other than the trivial G-AFM, where every pair of first neighbors is antiparallel. For instance, a second-neighbor AFM coupling can lead to a C-AFM order where chains of parallel-aligned moments order antiparallel (antiferromagnetically) between them (see Fig. \ref{fig:magstruct}) \cite{bacci1991spin,sushkov2001quantum}. 
If these chains are aligned along the basal plane, it is expected that below the AFM ordering temperature $T_{N}$, the lattice lowers its symmetry to orthorhombic \cite{rotter2006magnetoelastic}.
Although symmetry-conserving distortions at the AFM transition have been easily detected \cite{lindbaum2002chapter}, high-precision x-rays experiments (up-to $\vert a-b\vert/a \sim 2\times 10^{-4}$) do not show any difference between $a$ and $b$ lattice parameters \cite{rotter2006magnetoelastic,granado2006magnetic}.
Such an intriguing absence of lattice symmetry breaking at the N\'eel transition in Gd-based AFM systems has been referred to as the \textit{magnetoelastic paradox} \cite{rotter2006magnetoelastic}.

In this work we present very sensitive dilation experiments across the N\'eel transition in GdCoIn$_5$ ($T_{N}\approx$30 K)\cite{betancourth2015low} and GdRhIn$_5$ ($T_{N}\approx$40 K)\cite{pagliuso2001crystal}. 
Both systems show, for temperatures below $T_N$, an abrupt change of the longitudinal linear forced magnetostriction in an external field of $\sim 1$ Tesla.
We perform detailed calculations which indicate that the dipolar interaction, the crystal electric field, and the strain dependence of the magnetic exchange couplings are all essential to account for the observed lattice distortions and magnetic structure.
We also show that the dilation data is compatible with the existence of a tetragonal to orthorhombic distortion of the lattice at the N\'eel transition. The predicted orthorhombic distortions are below the x-ray resolution and result from the competition between the different magnetoelastic couplings.

The rest of this paper is organized as follows. Section \ref{SeccionExperimentos} presents the main experimental and theoretical results for the magnetostriction and thermal expansion data for the GdCoIn$_5$ and GdRhIn$_5$ compounds. Section \ref{sec:theory} presents the model, while Sec. \ref{sec:num} presents the numerical simulations and their interpretation. Finally in Sec. \ref{sec:conclusions} we present the conclusions.


\section{Main Results}\label{SeccionExperimentos}
\begin{figure}[h]
\includegraphics[width=0.5\columnwidth]{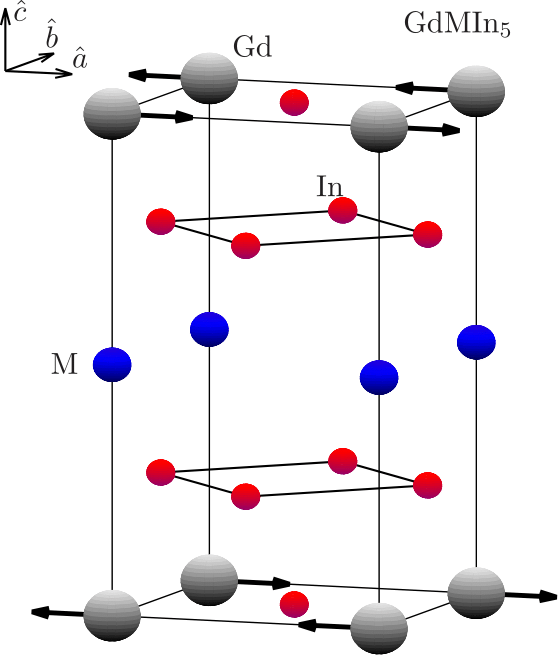}
\center
	\caption{(Color online). Crystal structure of the 115 compunds. The observed magnetic configuration (C-AFM) is indicated by the thick arrows on the Gd atoms. The spins are parallel to the $\hat{a}$-axis.} 
\label{fig:magstruct}
\end{figure}

We present below the main experimental and theoretical results on the magnetostriction and thermal expansion data for the GdCoIn$_5$  and GdRhIn$_5$ compounds. Both materials crystallize in the tetragonal HoCoGa$_5$ structure (see Fig. \ref{fig:magstruct}). 
The magnetic structure inferred by x-ray experiments in \GRI~is of C-type~\cite{granado2006magnetic} (see Fig. \ref{fig:magstruct}).

High quality single crystals of GdCoIn$_5$ (GCI) and GdRhIn$_5$ (GRI) were grown by the self-flux technique and characterized as described elsewherẹ~\cite{betancourth2015low}. 
The specific heat and the magnetic susceptibility data are compatible with $S=7/2$ spins at the Gd$^{3+}$ ions coupled by exchange interactions mediated by the conduction electrons. The main difference between \GCI~ and \GRI~ is the larger exchange coupling along the c-axis in the latter which leads to a higher \TN~\cite{facio2015co}.

Platelet-shaped crystals of typical size 1 $\times$ 1 $\times$ 0.4 mm$^3$ were selected for the dilation experiments which were performed with a high resolution ($\Delta L \leq$ 1 \AA) capacitive dilatometer~\cite{schmiedeshoff2006versatile}. 
All dilation experiments under magnetic field were carried out in the longitudinal configuration, i.e. with the magnetic field $B$ parallel to the sample dimension $L$ being measured.

\begin{figure}[h]
\includegraphics[width=\columnwidth]{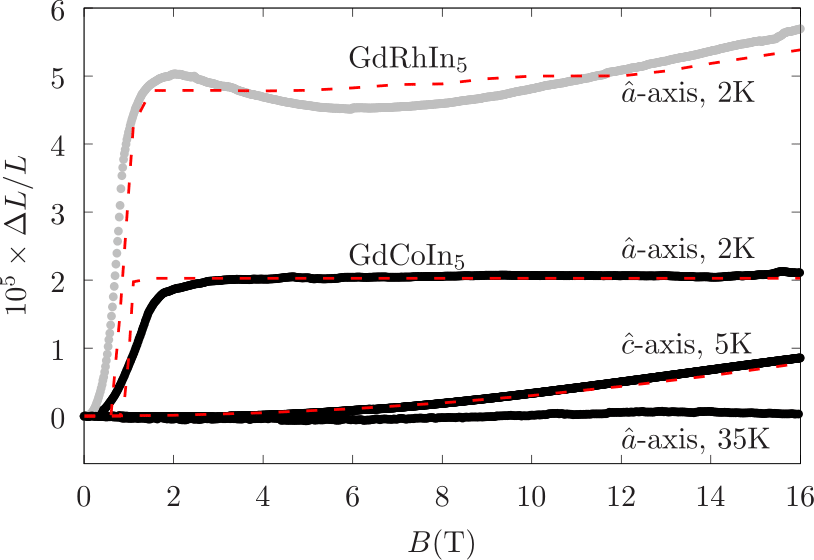}
\center
	\caption{(Color online) Solid symbols: Experimental forced magnetostriction for GdCoIn$_5$ (black symbols) and GdRhIn$_5$ (grey symbols) at different temperatures and along different directions as indicated in the figure. Dashed lines: calculted forced magnetostriction (see main text).}
\label{DLvsB}
\end{figure}

Figure \ref{DLvsB} summarizes the most important experimental observations of this work. It displays the forced magnetostriction (lattice dimension change driven by an external magnetic field) along the $\hat{a}$-axis for both GCI and GRI. At temperatures below $T_{N}$, $\Delta L_a/L_a$  shows a sudden increase at an in-plane field $B^\star \sim 1$ T. Above this field, $\Delta L_a/L_a$ becomes field independent for GCI and shows a weak increase with increasing magnetic field for GRI. No hysteresis effects are observed.  At temperatures above $T_{N}$ the magnetostriction becomes negligible and no sudden change is observed.
This type of change in $\Delta L$  for $T<T_N$ is usually seen on ferromagnetic materials  and attributed to the change from the zero magnetization to the saturated magnetic state (see chapter 1 on Ref. \cite{goran2000handbook}). 
As we will see in the following Section, in these antiferromagnetic 
systems an applied magnetic field induces
a spin-flop transition (at the field $B^\star$) to a metastable state where the spins point mainly along the $\hat{c}$-axis\cite{doi:10.1080/00018737700101433,johnston2017influence}. 
To account for the value $B^\star$ and the absence of hysteresis both the dipole interaction and the effect of the crystal field in second order perturbation theory needs to be considered in the calculations. 
The theoretical results (dashed lines in Fig. \ref{DLvsB}), were obtained using fitting interaction parameters constrained to the range of estimated values (see Sec. \ref{sec:theory}).

Along the $\hat{c}$-axis, the field dependence of the forced magnetostriction (MS) is quadratic at all temperatures as it is shown in Fig. \ref{DLvsB} for GCI. GRI mimics this behaviour (not shown here).

\begin{figure}[h]
\includegraphics[width=\columnwidth]{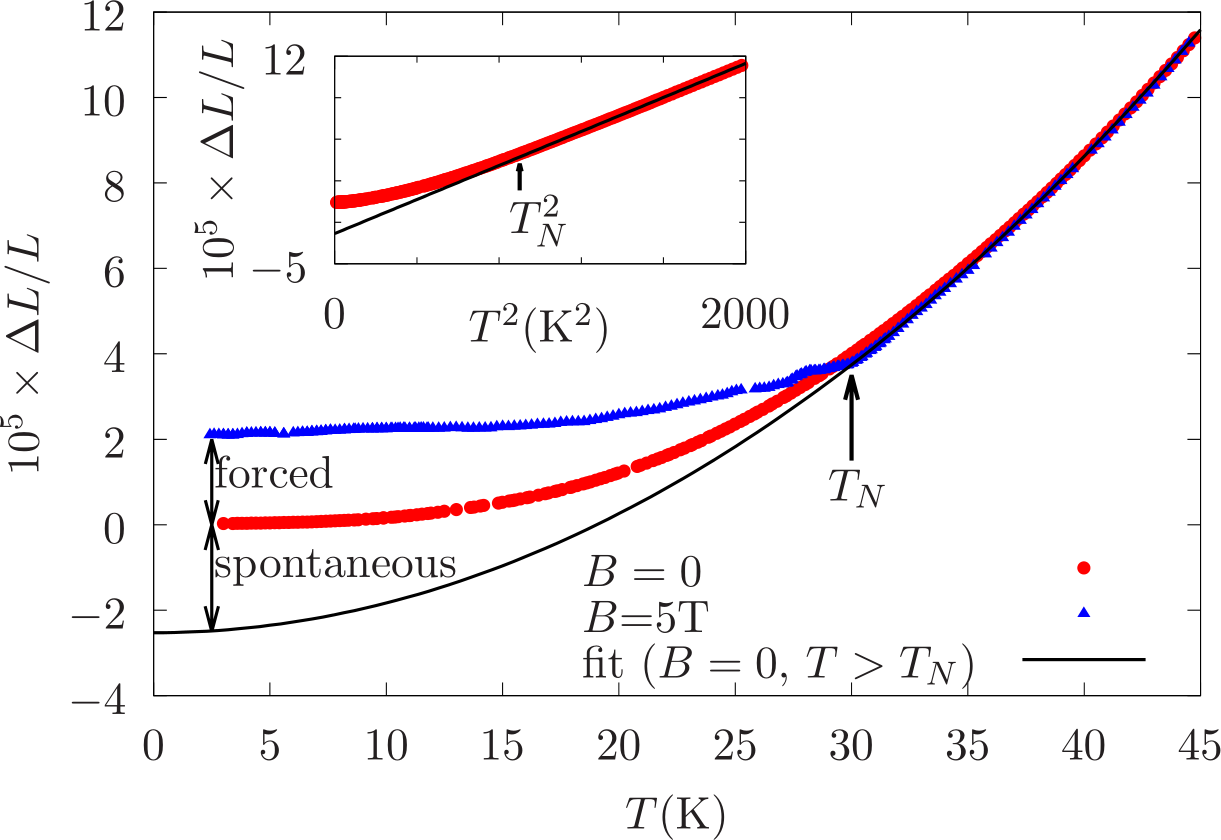}
\caption{(Color online) Thermal expansion along the $\hat{a}$-axis as a function of the temperature for GdCoIn$_5$ at $B=0$ and $B=5$ T.
The spontaneous and forced magnetostrictions are indicated in the figure. The non-magnetic contribution (thin line) is obtained from a linear fit of the thermal expansion versus $T^2$ at $B=0$ and for $T>T_N$ as shown in the inset.}
\label{DLvsT}
\end{figure}

The spontaneous and the forced magnetostrictions are a measure of the strength of the magnetoelastic couplings which can be extracted from the thermal expansion data.
The main panel of Fig. \ref{DLvsT} shows the $\hat{a}$-axis thermal expansion $\Delta L_a/L_a$ of GCI at zero field and at $B =$ 5 T $> B^*$. A small kink signposts the magnetic transition at $T_N$. 
In the paramagnetic state ($T > T_{N}$), $\Delta L_a/L_a \propto T^2$ as seen in the inset of Fig. \ref{DLvsT}. 
This quadratic non-magnetic thermal-expansion background is also included in the main panel and it extrapolates at low temperatures to $\Delta L_a(\text{non-magnetic},T\to 0)/L_a = -2.5 \times 10^{-5}$. 
We take the quadratic fit as the non-magnetic thermal expansion, and the difference between the zero field  thermal expansion curve and that fit corresponds to the spontaneous MS. Accordingly, the difference between the finite field and the zero field data is the forced MS. The total MS is the sum of the forced and the spontaneous MSs and can be obtained substracting to the finite field curve the non magnetic fit.

According to this analysis, the $\hat{a}$-axis spontaneous MS is positive in GCI, resulting in a zero temperature expansion 
$\Delta L_a(B=0,T\to 0)/L_a-\Delta L_a(\text{non-magnetic},T\to 0)/L_a  = 2.5 \times 10^{-5}$, while the forced MS is also positive giving 
$\Delta L_a(B=$5 T$,T\to 0)/L_a-\Delta L_a(B=0,T\to 0)/L_a= 2 \times 10^{-5}$. 
The zero-temperature spontaneous and forced MS's of both GCI and GRI are summarized in Table \ref{ComparingWithExperiments}.
An equivalent analysis can be performed from the thermal expansion data along the $\hat{c}$-axis. In this case, however, the spontaneous MS has the opposite sign $\Delta L_c(T\to 0)/L_c = -1.2 \times 10^{-5}$. 
\begin{table}[b]
	\caption{Measured and calculated spontaneous 
	and forced 
	magnetic expansions $\Delta L/L$ for $\hat{a}$-axis and $\hat{c}$-axis measurements. }
\label{ComparingWithExperiments}
\begin{tabular}{|c|c|c|c|}
\hline
	$10^{5}\times$ $\Delta L/L$ 		&   \GCI ($\hat{a}$) 	 & \GRI($\hat{a}$)  & \GCI ($\hat{c}$)  \\
\hline
Spontaneous (Exp.)						&	2.5			&	3.4 				&-1.2\\
Spontaneous (Calc.)						&	2.0			&	2.6 				&-4.7\\
Forced (Exp.)							&	2.0			&	4.9 				& 0.9\\
Forced (Calc.)							&	2.0			&	4.9 				& 0.9\\
\hline
\end{tabular}
\end{table}

The theoretical description of the MS data needs also to account for the pronounced anisotropy observed in the magnetic susceptibilities below \TN as it was reported previously \cite{betancourth2015evidence,pagliuso2001crystal}.
Interestingly, the observed difference between the $\hat{a}$-axis and $\hat{c}$-axis susceptibilities is rapidly suppressed as the magnetic field is raised above $\sim 1$ T \cite{betancourth2015evidence,van2007magnetic}.

\section{Magnetoelastic Hamiltonian} \label{sec:theory}

In this Section we present the model used to describe the experimental data. We present the different magnetic interaction terms in the Hamiltonian and analyze their coupling to the lattice degrees of freedom.
\subsection{Exchange interactions}
\begin{figure}
\includegraphics[width=0.6\columnwidth]{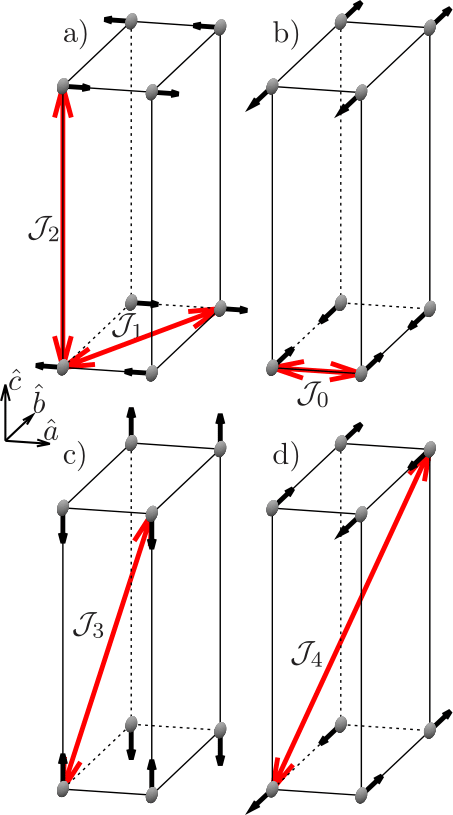}
	\caption{(Color online) Different magnetic arrangements and main exchange couplings. The small single headed arrows indicate the magnetic moment of the Gd$^{3+}$ ions, the double headed arrows indicate the exchange couplings $\mathcal{J}_i$ between the magnetic moments. a) Ferromagnetic chains along the $\hat{a}$-axis with the spins parallel to the same axis (C-AFM$_{aa})$. b) Ferromagnetic chains along the $\hat{a}$-axis with the spins parallel to the $\hat{b}$-axis (C-AFM$_{ab})$ and $\hat{c}$-axis c) C-AFM$_{ac}$. d) Analogous to a) with the chains and the spins along the $\hat{b}$-axis (C-AFM$_{bb}$). }
\label{Ordenamientos}
\end{figure}

The magnetic exchange interactions between localized magnetic moments at the Gd$^{3+}$ ions were determined by magnetic susceptibility and specific heat experiments combined with first principles calculations \cite{facio2015co,betancourth2015low}. The exchange interactions up to the fifth nearest neighbor are presented in Fig. \ref{Ordenamientos}. 
These couplings do not determine unambiguously the magnetic ground state as the energy only depends on the relative orientation of the magnetic moments.
). 
Figure \ref{Ordenamientos} presents four different magnetic moment arrangements  having the same exchange energy.

The exchange couplings $\mathcal{J}_{i}$, shown on Fig. \ref{Ordenamientos}, are modified when the lattice is distorted
\begin{align}
\mathcal{J}_{i} (\delta a , \delta b, \delta c ) &\simeq \mathcal{J}_{i} + \frac{d \mathcal{J}_{i}}{da} \delta a+ \frac{d \mathcal{J}_{i}}{d b} \delta b+ \frac{d \mathcal{J}_{i}}{d c} \delta c \\
\end{align}
where $\delta a$, $\delta b$, and $\delta c$ are uniform lattice distortions along the $\hat{a}$, $\hat{b}$, and $\hat{c}$ axis, respectively.
The magnetic exchange interaction between magnetic moments $\vec{S}_i$ at the Gd$^{3+}$ ions can be written as 
\begin{align} \label{EqSE}
	H_{E} &= \sum_{\langle i, j \rangle} \mathcal{J}_{i,j} (\delta a , \delta b , \delta c ) \vec{S}_i\cdot \vec{S}_{j}
\end{align}
where the $\mathcal{J}_{i,j}$ couplings are equal to $\mathcal{J}_0$ for first nearest neighbors, $\mathcal{J}_1$ for second nearest neighbors, etc.

Although this interaction and its dependence with the lattice distortions is the largest, it is not enough to explain the observed ground state configuration, nor the magnetoelastic data or the magnetic anisotropy. For example, the Hamiltonian given by Eq. (\ref{EqSE}) leads to the same energy for the C-AFM$_{aa}$ and the C-AFM$_{ac}$ configurations. As we show below, the dipolar interactions break this degeneracy.

\subsection{Dipolar interactions}
The dipolar interactions have an explicit dependence with the distance between spins:
\begin{align} \label{EqDip}
    H_{D}&=16.8\,\text{K}\sum_{i,j} \frac{a_{B}^3}{r_{ij}^3} \left(\vec{S}_i\cdot \vec{S}_j - \frac{3}{r_{ij}^2} (\vec{S}_i\cdot \vec{r}_i) (\vec{S}_j\cdot\vec{r}_{j})\right),
\end{align}
where K is Kelvin scale unit, $a_{B}$ is the Bohr radius, $r_i$ is the position of the $i$-th spin and $r_{ij}=|\vec{r}_i-\vec{r}_j|$
\footnote{The value of $16.8$K comes from the universal constants combination $\frac{g^{2} \mu_{0} \mu_{B}^{2} }{4 \pi a_{B}^{3}}$ where $g=2$ is the gyromagnetic factor for Gd$^{3+}$.}. 
Note that here and in what follows we take $k_B=1$ and use K for the energy units.
$H_D$ introduces a magnetic anisotropy. 
Since the distance between nearest-neighbor spins is larger along the $\hat{c}$-axis than along the $\hat{a}$ or $\hat{b}$ axes, 
the dipolar configurations with the lowest energy have 
the spins in the $a$-$b$ plane.
Still those states remain highly degenerate as a continuum of configurations with in-plane second nearest neighbors antiparallel have the same exchange and dipolar energies [see e.g. in Figs. \ref{Ordenamientos}a),  \ref{Ordenamientos}d), and  \ref{fig:confpath}].
The experimentally observed order are however the ones shown in Figs. \ref{Ordenamientos}a and \ref{Ordenamientos}d). 
The first state, Fig. \ref{Ordenamientos}a), becomes the lowest lying state under a distortion $a\to a + \delta a$ and $b\to b +\delta b$ such that the lattice parameters $a$ and $b$ become different (orthorhombic distortion). 

The uniaxial magnetic anisotropy along the $\hat{c}$-axis introduced by the dipolar interaction is however larger than what is inferred from the value of $B^\star$ (see Appendix \ref{app:minmod}). An additional source of magnetic anisotropy, which is due to crystal-field effects, needs to be considered to explain the value of $B^\star$.

\subsection{Crystal-field effects}
The Gd$^{3+}$ ion is, according to Hund's rules, in a 4f$^7$ state with $\mathcal{L}=0$.
The spin-orbit $\sum_{i}\vec{\sigma_{i}}.\vec{l_{i}}$ coupling, however, mixes this $\mathcal{L}=0$, $S=7/2$ state with a higher energy multiplet with $\mathcal{L}=1$, $S=5/2$ and $J=7/2$ (see Appendix \ref{app:gnds} for details), which is affected by the tetragonal crystal field.
As a consequence crystal field (CF) effects, although small, are present and need to be considered.

The total crystal-field effect can be written as
\begin{align} \label{EqCF}
	H_{CF}&=  \sum_{i} B_{2} S_{ic}^{2}  + A (\delta a - \delta b) [ S_{ia}^2 - S_{ib}^2 ],
\end{align}
where $S_{i\ell}$ is the component of the $i$-th spin along the $\hat{\ell}$-axis.
The first term in Eq. (\ref{EqCF}) is the intrinsic CF\cite{brooks1968spin} and the last term is induced by distortions between the $a$ and $b$ lattice parameters (see Ref. \cite{callen1965magnetostriction}).
The contribution due to $c$ deformations is negligible with respect to the intrinsic contribution $ B_{2} S_c^{2}$. The latter combined with the dipolar contribution to the anisotropy determine $B^\star$.

\subsection{Elastic energy}

The elastic energy for a uniform distortion can be approximated as
\begin{align} \label{EqEl}
    H_{el}&= \frac{1}{2} C^{ab}_{el}(\delta a^2+\delta b^2) + \frac{1}{2} C^{c}_{el} \delta c^2 
\end{align}
where the elastic constants $C^{ab}_{el}\sim C^{c}_{el} \sim C_{el}= 70000$K/\AA$^{2}$ are estimated from the elastic properties of materials of the same family of compounds~\cite{RSKumar,sarrao} and Density Functional Theory (DFT) results (see Appendix \ref{app:elc}).

\section{Numerical simulations}\label{sec:num}

The total Hamiltonian of the system is obtained combining Eqs. (\ref{EqSE}),  (\ref{EqDip}),  (\ref{EqCF}), and (\ref{EqEl}). 
\begin{align} \label{EqTotalEnergy}
H&=H_{el}+H_{D}+H_{E}+H_{CF}
\end{align} 
To evaluate this energy we approximate the large $S=7/2$ spins on the Gd$^{3+}$ ions with classical magnetic moments. We consider a lattice of $L\times L\times L$ sites. 
We find that $L\gtrsim12$ is enough to obtain $L$-independent results.

We minimize the energy considering uniform deformations $\delta a$, $\delta b$, $\delta c$, and spin rotations restricted to magnetic orders that preserve the 8-site magnetic unit cell of Fig. \ref{Ordenamientos}.

\begin{table}
	\caption{Coupling parameters. $\frac{d \mathcal{J}_0^a}{d a}$ corresponds to the rate of change of $\mathcal{J}_0$ due to a lattice distortion parallel to the coupling while $\frac{d \mathcal{J}_0^a}{d b}$ corresponds to a distortion in the perpendicular direction on the $a-b$ plane.
}\label{AllParameters}
\begin{tabular}{|c|c|c|}
\hline
Parameter										&   \GCI  & \GRI   \\
\hline
\hline
$\mathcal{J}_{0}$										&1.31 K	&1.21 K	\\
$\mathcal{J}_{1}$										&1.65 K	&1.74 K	\\
$\mathcal{J}_{2}$										&0.47 K	&1.43 K	\\
$\mathcal{J}_{3}$										&0.05 K	&-0.10 K	\\
$\mathcal{J}_{4}$										&-0.11 K	&-0.15 K	\\
\hline
$\frac{d \mathcal{J}_0^a}{d a}$& 0.61  K/\AA &	-2.4 K/\AA\\
$\frac{d \mathcal{J}_0^a}{d b}$& -1.39  K/\AA &	-2.9 K/\AA\\
$\frac{d \mathcal{J}_1}{d a}$ & 0.265	 K/\AA&	0.066	K/\AA\\
$A$												 & 0.49	K/\AA &	1.26 K/\AA	\\
	$B_{2}$										 &-0.058 K&-0.019 K	\\
\hline
$\frac{d \mathcal{J}_1}{d c}$							& -1.0 K/\AA&	--	\\
\hline
\end{tabular}
\end{table}

The coupling parameters used in the simulations of \GCI~and \GRI~are presented in Table \ref{AllParameters}. 
The exchange coupling parameters were obtained from DFT calculations \cite{facio2015co}. The remaining parameters were obtained fitting the magnetostriction and the magnetic susceptibility.  Reference values for the rate of change of the exchange coupling with the lattice parameter changes were obtained from DFT calculations~\footnote{See Supplemental Material at URL for details of the DFT calculations of the exchange couplings.}. Note that for simplicity only the variation of the largest exchange couplings ($\mathcal{J}_0$ and $\mathcal{J}_1$) with the lattice distortions was considered.

\subsection{Ground state for zero magnetic field ($B=0$)}
\begin{figure}
\includegraphics[width=0.8\columnwidth]{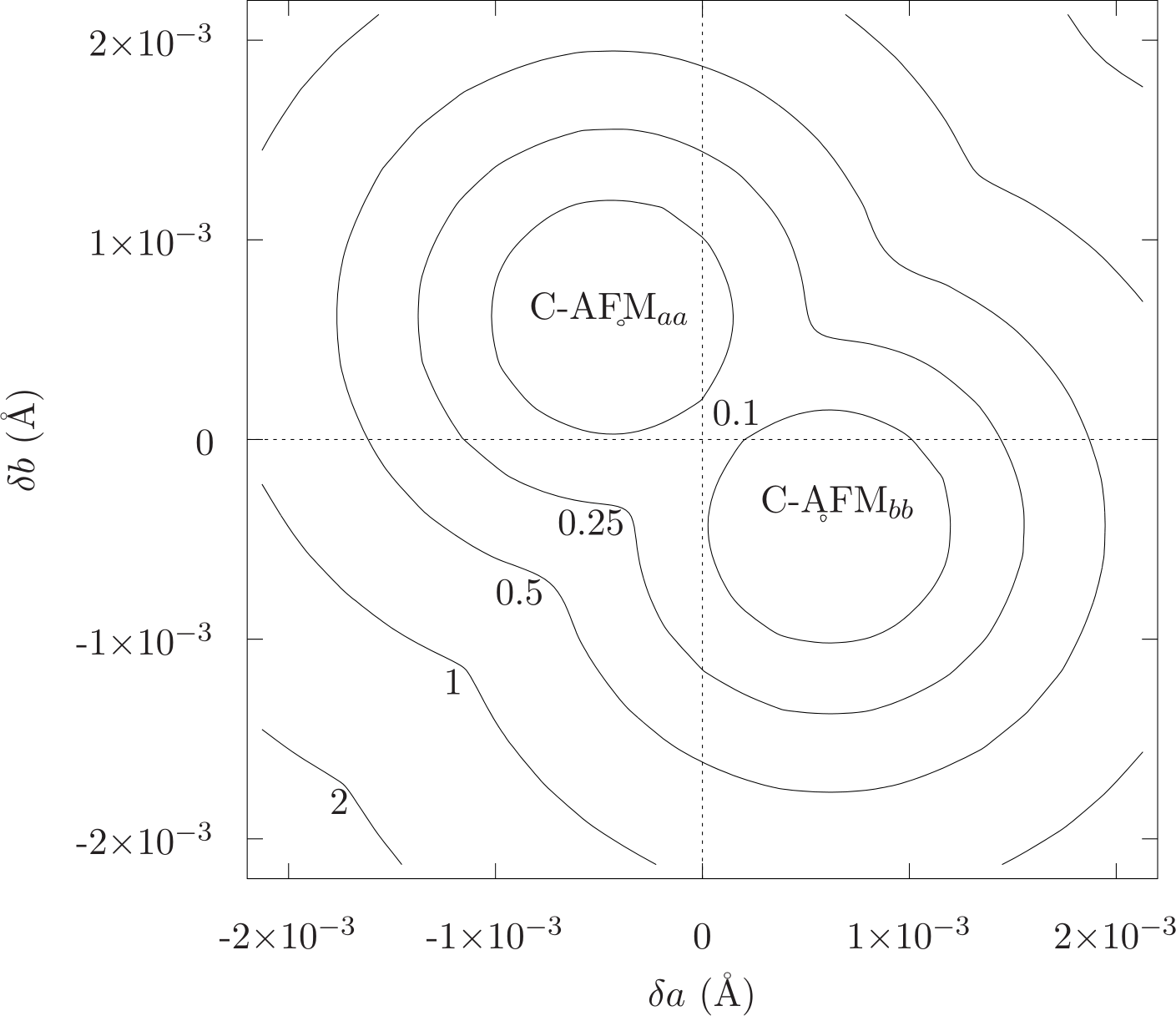}
	\caption{Total energy (in K) relative to the C-AFM$_{aa}$ configuration energy ($E_{aa}$) as a function of the changes in the lattice parameters $\delta a$ and $\delta b$.  }
\label{EnergyVsDeformationBZero}
\end{figure}

The total energy as function of $\delta a$ and $\delta b$ is shown on Fig. \ref{EnergyVsDeformationBZero}.
Two degenerate minima are obtained with magnetic orders C-AFM$_{aa}$ (see Fig. \ref{Ordenamientos}) and C-AFM$_{bb}$ which is related to C-AFM$_{aa}$ by a rotation of the lattice by \ang{90} around the $\hat{c}$-axis.
In the C-AFM$_{bb}$ configuration the spins and the spin chains are along the $\hat{b}$-axis. 
The distortions associated with these minima satisfy $\delta a_{aa} \ne \delta b_{aa}$ and  $\delta a_{aa} = \delta b_{bb}$, $\delta b_{aa} = \delta a_{bb}$.


The fact that $\delta a_{aa} \ne \delta b_{aa}$ indicates that the ground state crystal symmetry is reduced from tetragonal to orthorhombic. The magnitude of these distortions is however below the precision of high resolution XRD.

From here on we assume that the real system is composed by a mixture of these two states and consider the average of both distortions
\begin{align} \label{averagedeformations}
	\bar{\delta a} &= (\delta a_{aa} + \delta a_{bb})/2  \\ 
	\bar{\delta b} &= (\delta b_{aa} + \delta b_{bb})/2  
\end{align} 
to compare with the experimental results. Assuming a homogeneous distortion of the lattice we have $\Delta L_a/L_a = \bar{\delta a}/a$.  
In the zero field case ($B=0$) we have $\bar{\delta a}/a\equiv \bar{\delta b}/b$, which corresponds to the spontaneous MS, since the model does not consider non-magnetic distortions.
These distortions are presented in Table \ref{ModeledDistortionsOnPlane} and show a good agreement with the experimental results (see Table  \ref{ComparingWithExperiments}). 

\begin{table}[b]
	\caption{
		Calculated distortions $\delta a$, $\delta b$ for the $B=0$ and $B$=2 T (field parallel to the $\hat{a}$-axis) cases.
		The Lattice parameter $a$ is $4.568(3)$\AA~for \GCI~and 4.651(8)\AA~for \GRI~(see Ref. \cite{granado2006magnetic}).
		}
\label{ModeledDistortionsOnPlane}
\begin{tabular}{|c|c|c|c|}
\hline
										&B				&   \GCI 	 &  \GRI  \\
\hline	
\hline	
	$\delta a_{aa}$ & 0	&-4.4$\times 10^{-4}$\AA 		& -4.8$\times 10^{-4}$\AA	 \\
	$\delta a_{bb}$ & 0	&6.2$\times 10^{-4}$\AA 		& 7.2$\times 10^{-4}$\AA	 \\
$\delta b_{aa}$ & 0	&6.2$\times 10^{-4}$\AA 		& 7.2$\times 10^{-4}$\AA	   \\
$\delta b_{bb}$ & 0	&-4.4$\times 10^{-4}$\AA 		& -4.8$\times 10^{-4}$\AA	   \\
	$10^{5}\times \Delta L/L$		& 0		&	2.0 		&	  2.6	 \\ 
\hline
$\delta a_{ac}$&2 T	&-2.5 $\times 10^{-4}$\AA		&-0.2$\times 10^{-4}$\AA	\\
	$\delta a_{bb}$& 2 T		&6.2 $\times 10^{-4}$\AA		&7.2$\times 10^{-4}$\AA	\\
	$\delta b_{ac}$& 2 T		&4.4 $\times 10^{-4}$\AA		&1.3$\times 10^{-4}$\AA	  \\
	$\delta b_{bb}$ &2 T		&-4.4 $\times 10^{-4}$\AA		&-4.8$\times 10^{-4}$\AA	   \\
	$10^{5}\times \Delta L/L$ &2 T	&	4.0 	&	  7.5	 \\ 
\hline
\end{tabular}
\end{table}

\subsection{Magnetostriction}
To analyze the effect of an external magnetic field on the striction we include the Zeeman coupling
\begin{align} \label{EqZE}
H_{Z} &= \sum_{i} -g \mu_{B} \vec{S}_i\cdot \vec{B}.
\end{align} 
Under a magnetic field along the $\hat{a}$-axis, the energy of the C-AFM$_{aa}$ spin configuration remains unchanged while the energy of the C-AFM$_{bb}$ configuration [see Fig. \ref{Ordenamientos} d)] is reduced. 
A large energy barrier separates however these two configurations. To change from C-AFM$_{aa}$ to C-AFM$_{bb}$ at zero field,  the distortions need to be interchanged $\delta a \leftrightarrow \delta b$, the spins rotated $90^\circ$ interchanging nearest neighbour correlations from antiferromagnetic along the $\hat{a}$-axis to ferromagnetic, and vice-versa along the $\hat{b}$-axis. 
Although the C-AFM$_{ac}$ order has a larger energy than C-AFM$_{bb}$, it is much closer in configuration space to C-AFM$_{aa}$. The spin-spin correlations do not change and the distortions are only slightly modified as they are determined to a large extent by these correlations. The exchange coupling only depends on the relative orientation of the spins.
Our numerical results show that for a field $B>B^\star$ the configuration CAFM$_{ac}$ has a lower energy than CAFM$_{aa}$ and a transition between these two metaestable phases occurs.
The critical field $B^\star$ is determined by the dipolar energy and the intrinsic crystal field parametrized by $B_{2}$. For these compounds the CF reduces the critical field as it lowers the energy of the CAFM$_{ac}$ configuration. 
At $B=B^\star$ an energy barrier separates the C-AFM$_{bb}$ and the C-AFM$_{ac}$ configurations. 

In our scenario, the regions of the sample which at zero field were in the C-AFM$_{aa}$ configuration, with distortions $\delta a_{aa},\delta b_{aa}$ have, at $B=B^\star$, a sudden change to the C-AFM$_{ac}$ spin configuration with different distortions $\delta a_{ac}$ and $\delta b_{ac}$ (see Table \ref{ModeledDistortionsOnPlane}). The regions of the sample in the C-AFM$_{bb}$ configuration, however, remain in it and the field only tilts the spins slightly in its direction.
This leads to a forced MS at a field $B>B^\star$ given by $\Delta L_a (B,T=0) /L_a-\Delta L_a (B=0,T=0) /L_a =\left[\delta a_{ac}(B)+\delta a_{bb}(B)-\delta a_{aa}(B=0)+\delta a_{bb}(B=0)\right]/(2L_a)$, which is presented in Table \ref{ComparingWithExperiments} for \GCI~ and \GRI~at $B=5$ T.

For low  fields, $B<B^\star$,  only the deformations associated with the C-AFM$_{bb}$ order change. 
This is reflected on a small change of the total deformation. 
Around the critical field there is a sudden change of the deformations following the change of the magnetic order from C-AFM$_{aa}$ to CAFM$_{ac}$.


\begin{figure}
\includegraphics[width=1\columnwidth]{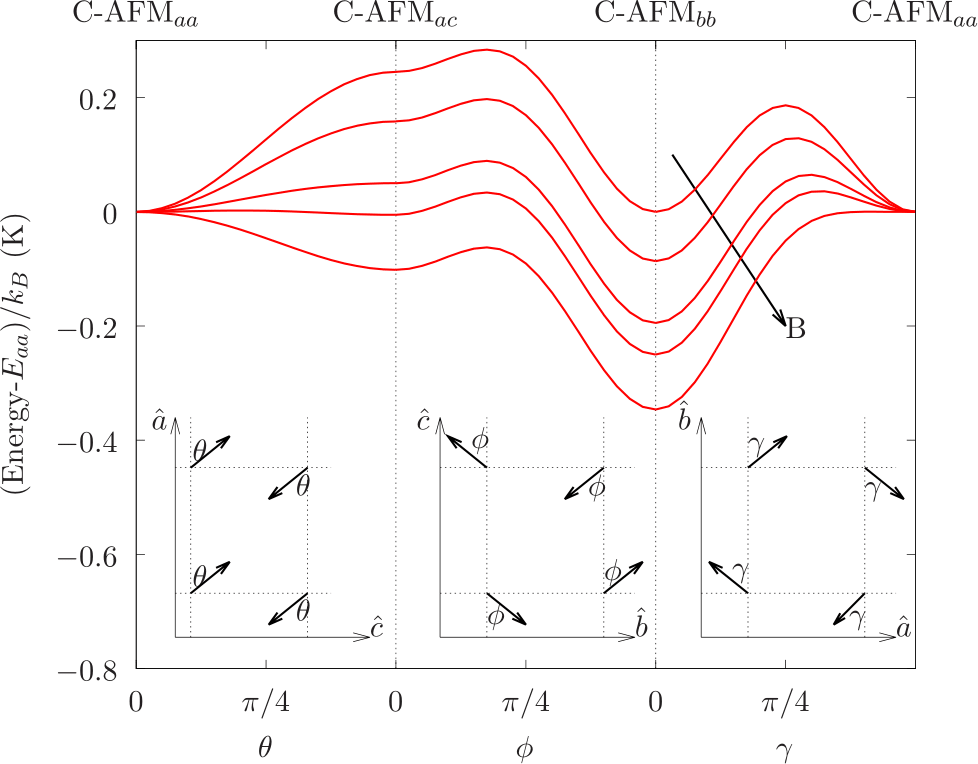}
	\caption{(Color online) Total energy of the system relative to the C-AFM$_{aa}$ configuration energy for a path in configuration space joining the C-AFM$_{aa}$, C-AFM$_{ac}$, and C-AFM$_{bb}$ phases. The spins are tilted uniformly by an angle as indicated in the figure.
	Different lines correspond to different values of the external magnetic field parallel to the $\hat{a}$-axis ($B$=0, 0.5T, 0.75T, 0.85T, 1T). }
\label{fig:confpath}
\end{figure}
Figure \ref{fig:confpath} presents the energy of the system for a path in the space of spin configurations that passes through the C-AFM$_{aa}$, C-AFM$_{ac}$, and C-AFM$_{bb}$ phases. To go from one configuration to the other, each spin is rotated by the same angle and the energy is minimized with respect to distortions $\delta a$, $\delta b$ and $\delta c$. In the absence of an external field, there are two degenerate minima for the C-AFM$_{aa}$ and C-AFM$_{bb}$ configurations. While the energy $E_{aa}$ of the C-AFM$_{aa}$ configuration does not depend on the magnetic field, both the C-AFM$_{ac}$ and C-AFM$_{bb}$ lower their energy with increasing magnetic field. At a field $B^\star \sim$ 0.85 T the C-AFM$_{aa}$ configuration becomes unstable and there is a transition to the C-AFM$_{ac}$ phase. This transition has a very small hysteresis loop of width $\sim 0.01$ T which is not observed in the experiments. Even for fields $B\sim 16$ T a barrier separates the C-AFM$_{ac}$ and C-AFM$_{bb}$ phases.

We present in Appendix \ref{app:minmod} a minimal model that captures the main physical ingredients and allows to obtain the model parameters from the experimental results.

The magnetostriction with the magnetic field parallel to the $\hat{c}$-axis is simpler to understand because, in this case, there is no spin-flop transition. As the magnetic field is increased, the local spins tilt in the $\hat{c}$ direction. This leads to a change in the spin-spin correlations of antiparallel spins 
\begin{equation}
    \delta \langle \vec{S}_i\cdot \vec{S}_j\rangle \sim -\frac{S^2}{2}\left( \frac{M(B)}{M_S} \right)^2 
\end{equation}
where $M(B)\ll M_S$ is the uniform magnetization along the $\hat{c}$-axis and $M_S$ is its saturation value.
Since $M(B)\propto B$, this results, to leading order, in a $c$ lattice parameter change $\delta c\propto B^2$.

\section{Summary and Conclusions} \label{sec:conclusions}
We analyzed the magnetoelastic properties of \GCI~and \GRI. We measured the thermal expansion and the longitudinal magnetostriction on single crystals of both compounds using a high resolution capacitive dilatometer and constructed from first principles a model to account for the observed data. 

These compounds present a number of intriguing properties. 
The observed magnetic order below $T_N$ is a C-type antiferromagnet which has a lower point symmetry than the lattice, but the expected tetragonal to orthorhombic distortion is not observed in high resolution XRD experiments. 
The in-plane dilation presents a sudden change at a field $B^\star\sim 1$ T and temperatures below the N\'eel transition which is not observed along the $\hat{c}$-axis (for fields along the same axis). 
Contrary to expectations for a spin-flop transition\cite{shapira1970magnetic,wang1994magnetic}, no hysteresis effects are observed at $B^\star$. 
Although crystal-field effects are expected to be negligible in the Hund's rule ground-state multiplet of the Gd$^{3+}$ ion, a magnetic anisotropy is clearly observed in these compounds.

To understand the observed magnetic structure and reproduce the magnetostriction and magnetic susceptibility data we find it necessary to consider the spin-spin exchange interactions and their dependence on the lattice distortions, the dipolar interactions, and the crystal-field effects due to the mixing of the terms $^8S$ and $^6P$ as a consequence of the spin-orbit coupling.

The exchange couplings and their dependence on lattice distortions were estimated from first principles DFT based calculations while the crystal-field model and parameters were obtained from second order perturbation theory. The final model parameters were obtained for each compound through a fitting procedure of the magnetostriction and magnetic susceptibility data. As a consistency check of the model parameters, we estimated the change in the N\'eel temperature with an applied hydrostatic pressure which shows and excellent agreement with the expected value from Ehrenfest's thermodynamic equations (see Appendix \ref{app:TN}).

Our model can fully account for the observed experimental data, including the observed spin-flop transition and the absence of evidence of tetragonal symmetry breaking in these compounds. The main assumption is that the magnetically ordered state is a spatially inhomogeneous mixture of the two possible degenerate ground states. This assumption is needed to explain the absence of asymmetry, in the magnetic susceptibility and magnetostriction, between the $\hat{a}$-axis and the $\hat{b}$-axis measurements.

Interestingly, there are other  examples in the literature of tetragonal Gd-based compounds that show a similar behaviour, i.e. antiferromagnetic order with magnetic anisotropy below $T_N$ and a sudden change of the forced magnetostriction under a moderate magnetic field: GdNi$_2$B$_2$C (Refs. \cite{rotter2006magnetoelastic,PhysRevB.77.134408}), GdAg$_2$ (Ref. [\onlinecite{Mehboob2011}]) and GdRu$_2$Si$_2$ (Ref. [\onlinecite{Devishvili2010}]) among others. 
The model discussed here could apply also to these cases.

\acknowledgments
We are supported by PIPs 112-201101-00832, 112-201501-00506, and 2015-0364 GI of CONICET and PICTs 2013-1045, 2016-0204, and 2015 0869 of the ANPCyT.



\appendix

\section{Simplified model}\label{app:minmod}
In this appendix we solve, at the mean field level, a simplified model that captures the main magnetoelastic properties of the system in the ordered phase. 
We focus the analysis on the 
modifications of
 the lattice parameters in the $a$-$b$ plane which present 
 sudden changes
  when a strong enough longitudinal magnetic field is applied. 
In the absence of an external magnetic field and assuming a C-AFM correlated state the model reads: 
\begin{align}
    \mathcal{F}&= \gamma_1 S_c^2+ \gamma_2(\delta a-\delta b)(S_a^2-S_b^2)\\&+\beta_1(\delta a-\delta b)+\beta_2(\delta a +\delta b)\\& + \frac{C}{2}\left( \delta a^2+\delta b^2 \right) -\lambda(S_a^2+S_b^2+S_c^2-1).
\end{align}
Here, the first two terms stem from the crystal field and the dipolar interaction. 
The third and fourth terms are due to the dependence of the exchange interactions on the lattice parameters. The fifth term is the elastic energy and in the last term, $\lambda$ is a Lagrange multiplier included to enforce the spin normalization $|\vec{S}|^2=1$. 
The parameters of the model are given by:
\begin{equation}
\gamma_1=\left(  \frac{\delta E_D}{S^2} + B_{2} \right) > 0 
\end{equation}
where $\delta E_D=E_D($C-AFM$_{ac})-E_D($C-AFM$_{aa})\sim 0.74$K, is the difference in dipolar energy between the C-AFM$_{ac}$ and  C-AFM$_{aa}$ phases. $\gamma_2=A$ is given by the variation of the crystal field with lattice distortions [see Eq. (\ref{EqCF})].
$\beta_1= ( \frac{dJ_0^a}{da} + \frac{dJ_0^a}{db} ) S^2$ and $\beta_2=\frac{dJ_1}{da} S^2$ are given by the variation of the nearest neighbor and in-plane diagonal magnetic couplings with the lattice distortions.

There are four sets of distortions and spin projections that satisfy $\partial\mathcal{F}/\partial \ell=0$ for all $\ell \in \{\delta a$, $\delta b$, $S_a$, $S_b$, $S_c$, $\lambda$\}:
\begin{widetext}
    \begin{table}[h] 
\begin{tabular}{|c|c|c|c|c|c|c|}
    \hline
    Magnetic order&$S_a$&$S_b$&$S_c$&$\delta a$&$\delta b$&Energy\\
    \hline
    \hline
    C-AFM$_{aa}$&1&0&0& $-\frac{1}{C}(\gamma_2+\beta_1+\beta_2)$&$\frac{1}{C}(\gamma_2+\beta_1-\beta_2)$&$-\frac{1}{2C}[(\gamma_2+\beta_1)^2+\beta_2^2]$\\
    C-AFM$_{bb}$&0&1&0& $\frac{1}{C}(\gamma_2+\beta_1-\beta_2)$&$-\frac{1}{C}(\gamma_2+\beta_1+\beta_2)$&$-\frac{1}{2C}[(\gamma_2+\beta_1)^2+\beta_2^2]$\\
    C-AFM$_{ac}$&0&0&1& $-\frac{1}{C}(\beta_1+\beta_2)$&$\frac{1}{C}(\beta_1-\beta_2)$&$\gamma_1-\frac{1}{2C}[\beta_1^2+\beta_2^2]$\\
    C-AFM$_{bc}$&0&0&1& $\frac{1}{C}(\beta_1-\beta_2)$&$-\frac{1}{C}(\beta_1+\beta_2)$&$\gamma_1-\frac{1}{2C}[\beta_1^2+\beta_2^2]$\\
    \hline
\end{tabular}
\end{table} 
\end{widetext}
The ground state is doubly degenerate (C-AFM$_{aa}$ and C-AFM$_{bb}$) while C-AFM$_{ac}$ and C-AFM$_{bc}$ are degenerate higher energy states since $\gamma_1>0$ and $\gamma_2\beta_1>0$.  C-AFM$_{ac}$ (C-AFM$_{bc}$) is however unstable with respect to a tilt of the spins, increasing $S_a$ ($S_b$)
. This fact preserves the spin-spin correlations (see Fig. \ref{fig:confpath} in the main text):  
\begin{equation}
    \left. \frac{d^2\mathcal{F}}{dS_a^2}\right|_{ac}=-\frac{4\beta_1\gamma_2}{C}-2\gamma_1<0
\end{equation}
A magnetic field along the $\hat{a}$-axis produces a magnetization $M\hat{a}$ in the states C-AFM$_{bb}$, C-AFM$_{ac}$ and C-AFM$_{bc}$ by tilting the spins along the same axis. This leads to a reduction of the energy of these states by $\sim B_a^2/T_N$ compared to the C-AFM$_{aa}$ state which remains unchanged.
At fields larger than
\begin{equation}
    B^\star\sim\sqrt{T_N}\sqrt{ \gamma_1+\frac{1}{2C}[\gamma_2^2+2\gamma_2 \beta_1]}\simeq \sqrt{T_N \gamma_1},
\end{equation}
the energy of the C-AFM$_{ac}$ state becomes lower than the energy of the C-AFM$_{aa}$ state.
Additionally the C-AFM$_{aa}$ state becomes unstable with respect to an increase in $S_c$: 
\begin{equation}
	\left. \frac{d^2\mathcal{F}}{dS_c^2}\right|_{ac}=\frac{4\beta_1\gamma_2}{C}+2\gamma_1-\frac{2 B_a^2}{T_N}\sim 2 \left[\left(B^\star\right)^2- B_a^2 \right]/T_N.
\end{equation}
This explains the transition between the metastable states C-AFM$_{aa}$ and C-AFM$_{ac}$ at $B=B^\star$.

\section{The ground state of Gd$^{3+}$}\label{app:gnds}

According to Hund rules, the state of maximum total angular momentum
projection $M=7/2$ of the ground-state multiplet  $^{8}S_{7/2}$ is

\begin{equation}
|0,7/2,7/2,7/2\rangle =\prod_{m=-3}^{3}f_{m\uparrow }^{\dagger
}|0\rangle ,  \label{0jj}
\end{equation}%
where $f_{l\sigma }^{\dagger }$ creates a 4f electron with orbital angular
momentum projection $l$ and spin $\sigma $. The notation of the states is $%
|\mathcal{L},\mathcal{S},J,M\rangle $, where  $\mathcal{L}$ ($\mathcal{S}$) is the total orbital angular momentum
(spin), and $J,M$ the total angular momentum and its projection. The other
states of the multiplet are obtained using repeatedly the lowering operator 

\begin{eqnarray}
J^{-} &=&\mathcal{L}^{-}+\mathcal{S}^{-},  \notag \\
\mathcal{L}^{-} &=&\sum\limits_{\sigma }\sum\limits_{m=-2}^{3}a(m)f_{m-1\sigma
}^{\dagger }f_{m\sigma },  \notag \\
\mathcal{S}^{-} &=&\sum\limits_{m=-3}^{3}f_{m\downarrow }^{\dagger }f_{m\uparrow }, 
\notag \\
a(m) &=&\sqrt{12-m(m-1)}.  \label{jm}
\end{eqnarray}%
The spin-orbit interaction

\begin{eqnarray}
H_{\lambda } = \lambda \sum\limits_{i}\mathcal{L}_{i}\cdot \mathcal{S}%
_{i} &=& \frac{\lambda}{2} [ \sum\limits_{m=-2}^{3}a(m)\left(f_{m\downarrow
}^{\dagger }f_{m-1\uparrow } + \text{H.c.}\right)    \notag \\
 &+& \sum\limits_{m} m\left(f_{m\uparrow}^{\dagger }f_{m\uparrow}-
 f_{m\downarrow}^{\dagger }f_{m\downarrow}\right)],  \label{hso}
\end{eqnarray}%
conserves the components of the total angular momentum $J$ but
modifies $\mathcal{L}$ and $\mathcal{S}$. 

We obtain 

\begin{eqnarray}
H_{\lambda }|0,7/2,7/2,M\rangle  &=&V|1,5/2,7/2,M\rangle ,  \notag \\
V &=& \sqrt{14} \lambda .  \label{vsoc}
\end{eqnarray}

The result is independent of $M$ as can be easily shown using the fact that $%
\mathbf{J}$ commutes with $H_{\lambda }$. We obtain the ground state
multiplet $|g,M\rangle $ of Gd$^{3+}$ solving a 2$\times $2 matrix (the same
for each $M)$

\begin{equation}
\left(\begin{array}{cc}
0 & V \\ 
V & E \\%
\end{array}\right)%
,  \label{matrixso}
\end{equation}%
where $E$ is the energy difference between the multiplets $^{6}P_{7/2}$ and  
$^{8}S_{7/2}$ for $\lambda =0$. 
Then, from the lowest lying state of Eq. (\ref{matrixso})
we obtain with $u,v>0$ and $u^{2}=1-v^{2}$

\begin{eqnarray}
|g,M\rangle  &=&u|0,7/2,7/2,M\rangle -v|1,5/2,7/2,M\rangle ,  \notag \\
v^{2} &=&\frac{1}{2}-\frac{E}{4\sqrt{(E/2)^{2}+V^{2}}}.  \label{gm}
,  \label{gstate}
\end{eqnarray}%
From optical experiments (Fig. 8 of Ref. \cite{dieke}) we estimate $E\simeq
32000$ cm$^{-1}$. From the same reference, averaging the total spin-orbit
splitting $\Delta =J_{\max }(J_{\max }+1)\lambda /(4S)$ between $J_{\max }=6$
and $J_{\min }=0$ for the $^{7}F$ terms ($\mathcal{L}=\mathcal{S}=3$) of Eu$^{3+}$ (configuration 4f$%
^{6}$) and Tb$^{3+}$ (configuration 4f$^{8}$), we estimate  $\Delta =5500$ cm$^{-1}$, 
which implies $\lambda \simeq 1571$ cm$^{-1} \simeq 0.19$ eV. 
This gives $v^{2}=0.0307$.
The value of $\lambda$ is similar to $\lambda \simeq 1508$ cm$^{-1}$ 
reported by Carnall {\it et al.} for Gd doped LaF$_3$ \cite{car}.

\section{Estimation of the elastic constants} \label{app:elc}
To obtain estimations for the elastic constants of \GRI~and \GCI~we used a combitation of experimental results for related materials and density functional theory (DFT) calculations.

High pressure x-ray diffraction (XRD) experiments on CeRhIn$_5$ and CeCoIn$_5$ report $K\sim 78$GPa for the bulk modulus of both materials \cite{RSKumar}.
This leads to an elastic energy per atom
\begin{equation}
	E_{el}(\delta a,\delta b, \delta c)= \frac{1}{2} E\left[  c \left(\frac{\delta a}{a}\right)^{2} +  c \left(\frac{\delta b}{b}\right)^{2} + \frac{a^2}{c} \left(\frac{\delta c}{c}\right)^{2} \right].  \nonumber
\end{equation}
Here $E=3 K (1 - 2 \nu)$ is the Young modulus assuming an isotropic material where $\nu$ is Poisson's ratio.
This results in anisotropic elastic constants along the $\hat{a}-\hat{b}$ and $\hat{c}$ axes, $C_{el}^{ab}=  E  c $ and $C_{el}^{c}=  E  a^{2}/c $, respectively.
Using $\nu\sim 0.22$ (see below), we obtain $C_{el}^{a} \sim 71000$K/\AA$^{2}$ and $C_{el}^{c} \sim 27000$K/\AA$^{2}$. 

\begin{figure}
\includegraphics[width=\columnwidth]{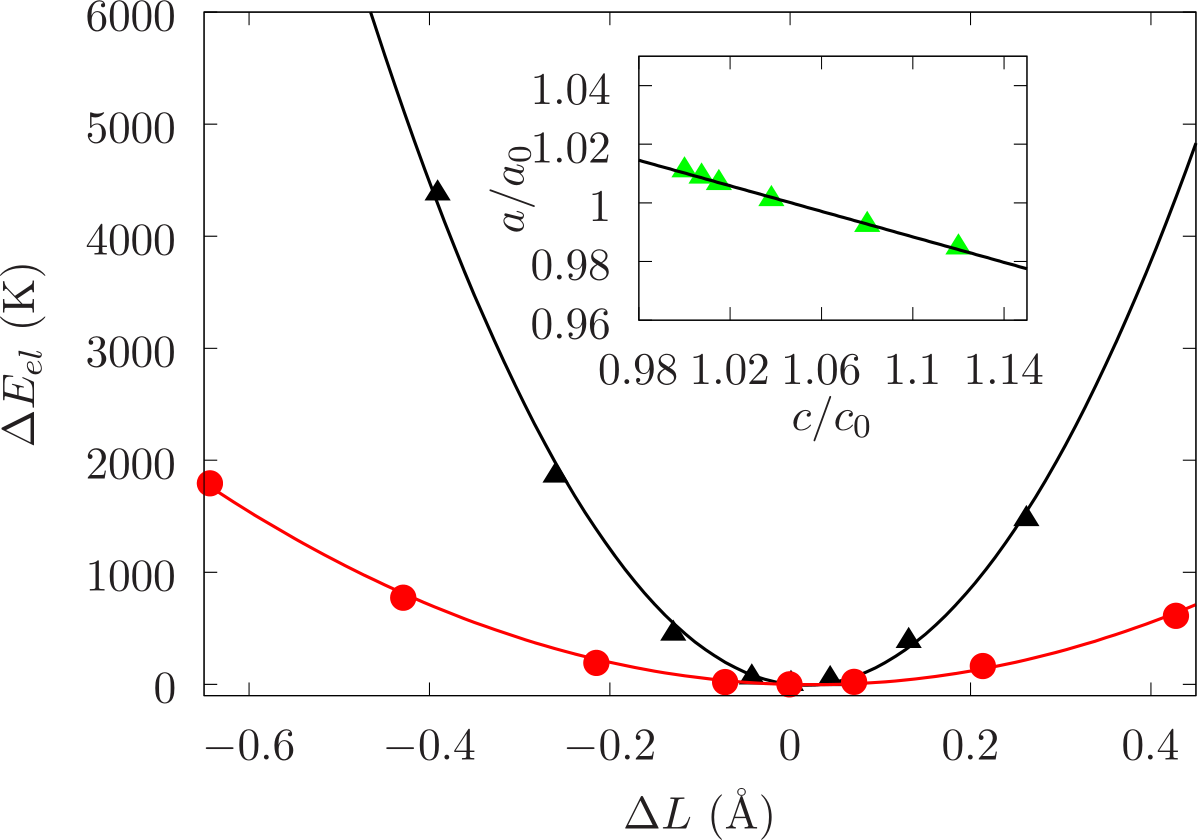}
	\caption{(Color online) Total energy versus $\delta a=\delta b=\Delta L$ for fixed $c=c_0$ (black triangles) and versus $\delta c=\Delta L$  for fixed $a=b=a_0$ (red disks). The lines are quadratic fits. Inset: Deformation on $a$ lattice parameter due to a change on $c$, relative to the equilibrium values $a_0$ and $c_0$, respectively. The line is a linear fit used to extract the Poisson's ratio.}
\label{ModuloYoungPoissonDFT}
\end{figure}

The DFT calculations were performed using a supercell of $2\times 2\times 2$ unit cells.
On Fig. \ref{ModuloYoungPoissonDFT} we show the change on the total energy as one of the unit cell lattice parameters ($a$ or $c$) is changed.
The quadratic behaviour allows us to obtain the young modulus in both situations resulting
\begin{equation}
	C_{el}^{a}=52000 \text{K/\AA}^{2} ; 	C_{el}^{c}=8000 \text{K/\AA}^{2}. \nonumber
\end{equation}
We have also computed the Poisson's ratio [see Fig. \ref{ModuloYoungPoissonDFT}b)] resulting in $\nu\sim 0.22$.


\section{Predicted effect of pressure on $T_{N}$} \label{app:TN}
As a consistency check of the model parameters we calculate here the change of T$_{N}$ with external pressure.
We find a good agreement between the model results and the inferred from Ehrenfest's thermodinamic equations (approximately $\sim 0.8$K/GPa and $\sim 1.2 $K/GPa respectively, see below).

An external hidrostatic pressure $P$ produces changes $\delta b/b=\delta a/a=\delta c/c= -P/E$ on the lattice parameters. 
This modifies the N\'eel temperature for the C-AFM order
\begin{align}
T_{N}=\frac{J(J+1)}{3k_{B}} (4\mathcal{J}_{1}+2\mathcal{J}_{2}-8\mathcal{J}_{4}),
\end{align}
through the exchange coupling constants $\mathcal{J}_i$ dependence on the lattice parameters.

For our model parameters (Table \ref{AllParameters}), we find that the changes in $T_N$ are dominated by the dependence of $\mathcal{J}_1$ on the lattice parameters
\begin{align}
    \delta T_{N} &\sim-4 P \frac{J(J+1)}{3k_{B}} (2 a \frac{1}{E}\frac{d\mathcal{J}_{1}}{da}  +c \frac{1}{E} \frac{d\mathcal{J}_{1}}{dc}  ) \\ \nonumber
\end{align}	

For an isotropic Young modulus of 131GPa (see appendix \ref{app:elc}), we obtain
\begin{align}
    \frac{\delta T_{N}}{P} \sim 0.8 \text{K/GPa}.
\end{align}


Ehrenfest's equations allow to relate the specific heat and the lattices changes on a second order transition
\begin{align}
	{dT_{N}} \over {dP} &= V_{m} T_{N} { {\Delta \alpha_{V}} \over {\Delta c_P}}.
\end{align}
For \GCI, using $V_{m}= 155.85$\AA$^{3}$, the lattice parameters of Table \ref{ModeledDistortionsOnPlane}, and the changes of specific heat and dilation reported in \cite{betancourth2015low}, we obtain
\begin{align}
	 \Delta \alpha_{V} \sim 6\times 10^{-6} /K ,\quad \Delta c_P \sim 15 \frac{J}{mol K},
\end{align}
which results in
\begin{align}
    \frac{\Delta T_{N}}{dP}	&\sim 1.2 \text{K/GPa}.
\end{align}


\bibliographystyle{apsrev4-1}

\end{document}